\documentclass{cimento}
\usepackage{graphicx}  

\title{Preliminary results of the analysis of the BATSE TTE data}
\author{
I.~Horv\'ath\from{ins:bolyai},
J.~P.~Norris\from{ins:gsfc},
J.~D.~Scargle\from{ins:ames}
   \atque
L.~G.~Bal\'azs\from{ins:konkoly}
}

\instlist{ 
\inst{ins:bolyai} Department of Physics, Bolyai Military Univ. 
H-1456 Budapest, POB 12,  Hungary 
\inst{ins:gsfc}GLAST Science Support Center, Code 661, 
NASA Goddard Space Flight Center, Greenbelt, MD 20771 
\inst{ins:ames}Space Science Division, MS 245-3, 
NASA Ames Research Center, Moffett Field, CA 94035-1000
\inst{ins:konkoly} Konkoly Observatory, H-1525 Budapest, POB 67, Hungary

}
\PACSes{\PACSit{98.70}{98.80}}
\begin{document}

\maketitle

\begin{abstract}
 The Compton Gamma Ray Observatory (CGRO) observed many types of 
data and one of them is the time-tagged photon events (TTE data). We use 
the Bayesian block analysis, using Bayesian statistics, analyses the 
TTE data.  Our results; calculations of duration (T100), count rates 
(burst photon numbers in different channels) and count peaks 
(in 64, 16 and 4 ms). We present the duration,  the peak duration 
  and the distance between peaks
  distributions. Principal Component Analysis (PCA) has been
also applied.
 The PCA shows interesting results, such as channel 4 
(highest energy channel) probably is very important.
\end{abstract}

\section{Introduction}

 The CGRO/BATSE observed many types of data about gamma-ray bursts (GRBs), and their statistical
analyses gave several useful results. For example, the 
 logN-LogS analyses (\cite{ref:mm}, \cite{ref:hmm})
gave useful information about the spatial distribution
of GRBs. The study of the time behavior of the spectra (\cite{ref:ry1},
\cite{ref:ry2}) led to the better understanding of the time dependence of GRBs.
  In this paper the time-tagged photon events (TTE) will be studied. 
The TTE 
data recorded the arrival time (within a two microsecond bin), energy 
(within four discriminator channels) and detector of each photon. 
The Bayesian block analysis, using Bayesian statistics, analyses the 
TTE data and the output is the most probable segmentation of the observation 
into time intervals during which the photon arrival rate is perceptibly 
constant.

In the BATSE database there are 532 burst TTE (time tagged event) data.
The TTE data contains the detection time as many as 32 768
 photons with a 2 microsecond time resolution in four energy channels. 
Many cases there were more than 32 768 photons during the burst time or 
there were burst's photons before the starting time of TTE.
We used the 273 bursts' TTE data, which were complete (covered the whole burst).

\begin{figure}
\includegraphics[width= {0.9\columnwidth}, angle=0]{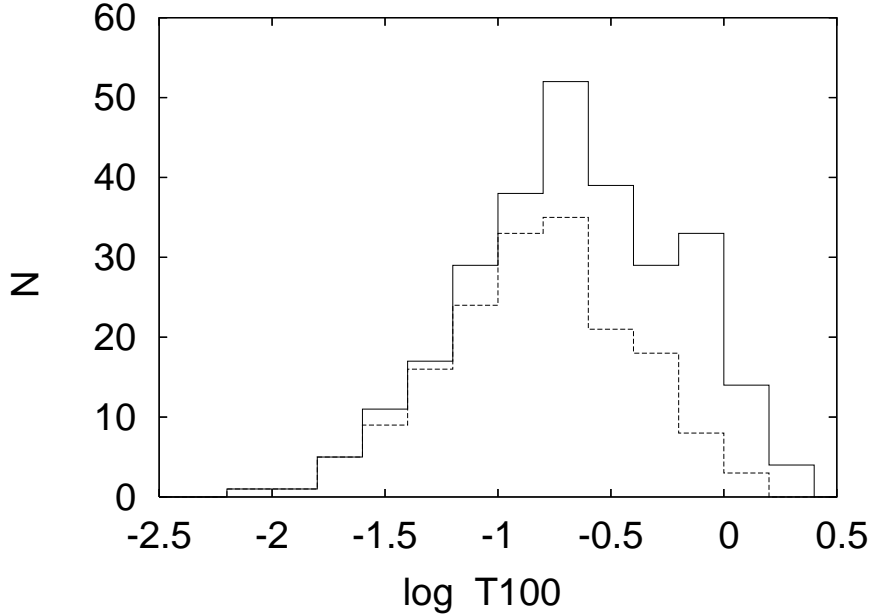}
\caption{The duration distribution of the analyzed 273 bursts (solid line)
and the duration distribution of the 174 one-peak bursts (dashed line).}
\end{figure}

\section{The Bayesian block analysis}

The Bayesian block (BB) analysis has been already developed in the 
literature for analyzing different data types 
\cite{ref:lor}, \cite{ref:Sivia}, \cite{ref:Scargle}, \cite{ref:Norris}. 
It is a method to find optimal changepoints
(times at which the count rate is modeled as
abruptly changing).  The marginal posterior
probability of the model is:

\begin{equation}
 L(M_i,D) = \int P(D / \Theta _i M_i) P(\Theta _i / M_i) d \Theta _i 
\label{eq:i}
\end{equation} 

where $M_i$ refers to the parameters specifying
the changepoint locations, and all other
parameters -- specifying the photon rates -- 
represented by $\Theta_i$ are marginalized as 
indicated by the integration in the equation.
The explicit form of this posterior for a block
of data depends on only two {\it sufficient
statistics}, namely the number of photons in
the block, and the length (in time) of the block.
The algorithm in 
\cite{ref:js}
yields the optimal block segmentation of the TTE
data.

\begin{figure}
\includegraphics[width= {0.90\columnwidth}, angle=0]{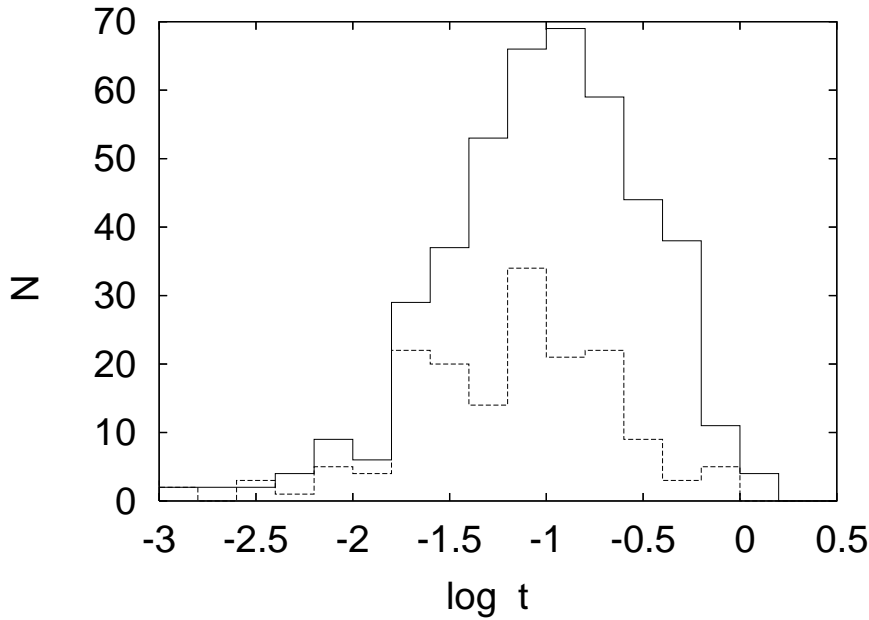}
\caption{The peak length distribution of the 431 well-defined peaks (solid line)
and the distance distribution between these peaks (dashed line).}
\end{figure}

\section{Calculation of the burst parameters}
  
Once one has the BB representation of the burst, one can calculate the burst parameters. 
First we have T100 rather than T90, since BBs show us the start and the end points of the burst. 
Second we need a background counts. Firstly we adjust the TTE data with 
cat64ms data. Secondly two 0,64 second long intervals were chosen. 
One before the burst and one after the burst. 
In other words ten consecutive bins were chosen before the burst 
and also after the burst from the cat64ms data for calculating the background. 
Third we assume during the burst the background rate was constant. 
After these one can calculate similar burst parameters then the BATSE catalog contains. 
One can calculate the counts in the four channels (like fluence) 
and also find the highest bin in different timescales. 
This can be called peakcounts (like peakflux). 
This was done in three timescales 64ms, 16ms and 4ms.
The natural definition of pulse width using
a block representation might be called  T100 rather than T90. 
{\bf Figure 1.} shows the duration distribution of the 273 bursts 
and the duration distribution of the 174 one-peak bursts. 
For peaks one can calculate the peak width or length rather than FWHM. 
{\bf Figure 2.} shows the peak length and the distance between peaks distributions.

\begin{table}
\caption{The eigenvalues of the principal component analysis of the 
8 quantities of Gamma-Ray Bursts (T100, four count rates, three count peaks). 
The first three PCs are important and the cumulative percentage is 
96\%, which means only three variables can explain 
96\% of the whole information.}
$$
\begin{array}{llll}
\hline
 Principal &         &     \%  \  of  & Comulative \\
 Component & Eigenvalues & Variance &  percentage   \\
\hline
\textbf{1} & \textbf{5.299} & \textbf{66.23} &  \textbf{66.23} \\
\textbf{2} & \textbf{1.723} & \textbf{21.54} &  \textbf{87.77}  \\
\textbf{3} & \textbf{.676} & \textbf{8.45} &  \textbf{96.22 }\\
4 & .119 & 1.50 & 97.71 \\
5 & .070 & .88 & 98.6 \\
6 & .066 & .82 &  99.4 \\
7 & .027 & .34   &  99.74 \\
8 & .011 & .24 &  100.0 \\
 \hline
   \end{array}
$$
\end{table}

\section{Principal Component Analysis (PCA)}

Using the logarithm of these 8 parameters one can make a PCA. 
The Principal Component Analysis can show us which parameters are important 
to characterize the bursts. {\bf Table 1.} shows the PCA eigenvalues 
and {\bf Table 2.} shows the eigenvectors. The first PC is the sum of the all parameters. The second PC is mainly the difference between duration and peakcounts. The third PC is mostly channel 4 just itself.

\begin{table}
\caption{The eigenvectors of the principal component analysis of the 
8 quantities of Gamma-Ray Bursts (T100, four count rates (Ch1-4), 
three count peaks (P64, P16, P4)). }
$$
\begin{array}{lllllllll}
\hline
Eigenvectors & lg T100 & lg Ch1 & lg Ch2 & lg Ch3 & lg Ch4 & lg P64 & lg P16 & lg P4 \\
\hline
1  & .54 & .88 & .92 & .93 & .7 & .91 & .8 & .75  \\
2  & .8 & .29 & .27 & .26 & .12 & -.35 & -.58 & -.62   \\
3  &  -.07 & -.32 & -.21 & .16  & .7 & -.06 & -.04 & -.05 \\

 \hline
   \end{array}
$$
\end{table}

However our analysis use different timescale and only for the short bursts
 these results are a good agreement with  \cite{ref:bag}.
This does not means short and long bursts are similar. 
Our result meaning is the same 2-3 parameters can describe the 
BATSE observed parameters for all bursts. 
But the short and the long ones can be in different place in this 3D space.

\acknowledgments

Thanks are due to the valuable discussions with 
Z. Bagoly,  P. M\'esz\'aros
and G. Tusn\'ady.  This research was supported through OTKA grants T034549
and T48870.


\begin{thebibliography}{0}

\bibitem{ref:bag} \BY{Bagoly, Z., Mesz\'aros, A., Horv\'ath, I., 
Bal\'azs, L.G. \& Mész\'aros, P.} \IN{ApJ}{498}{1998}{342.} 

\bibitem{ref:hmm} \BY{Horv\'ath I., M\'esz.\'aros, P. \atque 
M\'esz\'aros A.} \IN{ApJ} {470} {1996} {56.}

\bibitem{ref:js} \BY{Jackson~B. \etal} \IN{IEEE Signal Processing Letters}{12}
{2004}{105.} 

\bibitem{ref:lor} \BY{Loredo~T.J.} 
\IN{Statistical Challenges in Modern Astronomy, ed. Feigelson and Babu,  }
{}{Springer, New York}{1992, p. 275.} 

\bibitem{ref:mm} \BY{M\'esz\'aros A. \atque  M\'esz\'aros P.} \IN{ApJ} 
{466} {1996} {29.}

\bibitem{ref:Norris} \BY{Norris, J.P. Scargle, J.D. Bonnell, J.T. } 
\IN{Gamma-Ray Bursts in the Afterglow Era, Proc. of the Int. workshop held in Rome.
Ed. by Costa, Frontera, and Hjorth.}{}
{  Springer, Berlin}{2001, p. 40.} 

\bibitem{ref:ry1} \BY{Ryde~F. \etal} \IN{A\&A}{in press}{2005}{astro-ph/0411219}
\bibitem{ref:ry2} \BY{Ryde~F.} \IN{A\&A}{429}{2005}{869.}

\bibitem{ref:Scargle} \BY{Scargle~J.D.} \IN{ApJ}{504}{1998}{405.} 
\bibitem{ref:Sivia} \BY{Sivia~D.S.} \IN{Data Analysis: A Bayesian Tutorial 
}{}{Clarendon, Oxford}{1996} 

\end{thebibliography}
\end{document}